\documentclass[submission,copyright,creativecommons]{eptcs} 
\usepackage{graphicx}
\usepackage{xspace}
\usepackage{url}

\providecommand{\event}{FTSCS 2012} 
\usepackage{breakurl}             

\title{Towards system development methodologies:\\ From software to cyber-physical domain}
\author{Maria Spichkova  \qquad\qquad  Alarico Campetelli
\institute{Institut f{\"u}r Informatik \\Technische Universit{\"a}t M{\"u}nchen}
\email{\quad spichkov@in.tum.de 
 \quad\qquad campetel@in.tum.de}
}
\def\titlerunning{From software to cyber-physical domain}
\def\authorrunning{M. Spichkova, A. Campetelli}
\begin{document}
\maketitle

\newcommand{\Focus}{\textsc{Focus}\xspace}

\begin{abstract}
In many cases, it is more profitable to apply existing methodologies than to develop new ones. This holds, especially, for  system development within the cyber-physical domain: until a certain abstraction level we can (re)use 
 the methodologies for the software system development to benefit from the  advantages of these techniques. 

\end{abstract}

\section{Introduction}
A lot of new ideas are just well-forgotten old ones, and a lot of newly developed methodologies are, in fact, the reinvention of the wheel.  
In many cases  new methods or languages are introduced to deal with cyber-physical systems, where the already existing techniques could be more suitable to represent them, especially after  an extension or adaptation to cover some special domain features.
Moreover, the existing techniques could provide a connection to the representation of systems from other domains. 

Cyber-physical systems 
are widespread in safety-critical domains, as vehicles, machines, aircrafts or medical instruments. A failure in these systems may lead to considerable loss of money or even endanger human lives. Therefore, it follows the importance of a correct behaviour of the systems, which can be guaranteed through analysis techniques, mostly in form of formal verification (cf.~\cite{TUM-I1008}). The support for a formal analysis approach is facilitated by a formal model representation, and a suitable modelling theory for these systems helps in their development, maintenance, simulation, and verification. 

The main challenge here is to combine two worlds, the physical and the virtual one:  software components operate in discrete program steps, meanwhile the physical components evolves over time intervals following physical constraints. 
Nevertheless, many physical properties can be represented in a similar manner as the software properties, e.g., in many cases it is possible to switch from 
the  continuous time representation to the digital one without loosing the essential properties of the represented system~\cite{DigitalClocks}. 

From our experience within a number of industrial projects, 
speaking about the system architecture and properties on a certain abstraction level, ones do not need to distinguish physical signals and component from the software ones. 
In fact this difference does not have any advantages and, moreover, could make the system description too complicated and hardly readable. Thus, until we are speaking about logical level, we can benefit from using the software system development processes. 
For this reason, we present a system development methodology, which is a generalization of two methodologies successfully evaluated on three case studies from the automotive domain, with a suggestion to apply its general ideas for the development of cyber-physical systems.

\vspace{-2mm}
\section{System Development Methodology}
One of the typical examples of a cyber-physical system from the automotive field is a Cruise Control System.   
We modeled two different variants of the system using two development methodologies with similar strategies but  different focal points.
The first case study~\cite{dentum_tb}, developing an Adaptive Cruise Control  
system with Pre-Crash Safety functionality,  was motivated and supported by DENSO Corporation, while the second  case study \cite{VerisoftXT_FMDS,spichkova_st2011}, developing a Cruise Control System  with focus on system architecture and verification, was supported by Robert Bosch GmbH. 
Another case study~\cite{dentum_tb_2}  motivated and supported by DENSO Corporation was the development of the KeylessEntry-System which was  not only a  comfort system with the distributed deployment, but also a system having a huge state space.   
Thus, a sample-property of  a Cruise Control System can be represented as follows: 
\emph{If the driver pushes the ACC-button while the system is On and none of the  switch-off constraints occurs, the system must accelerate the vehicle during the next time unit respectively to  the predefined acceleration schema.} 
This means that the system must analyze the information from sensors to check whether any switch-off constraints occurs, i.e. if the battery voltage is too low or if the gas pedal sensor fails.

\begin{figure}[ht!]
  \begin{center}
   \includegraphics[scale=0.45]{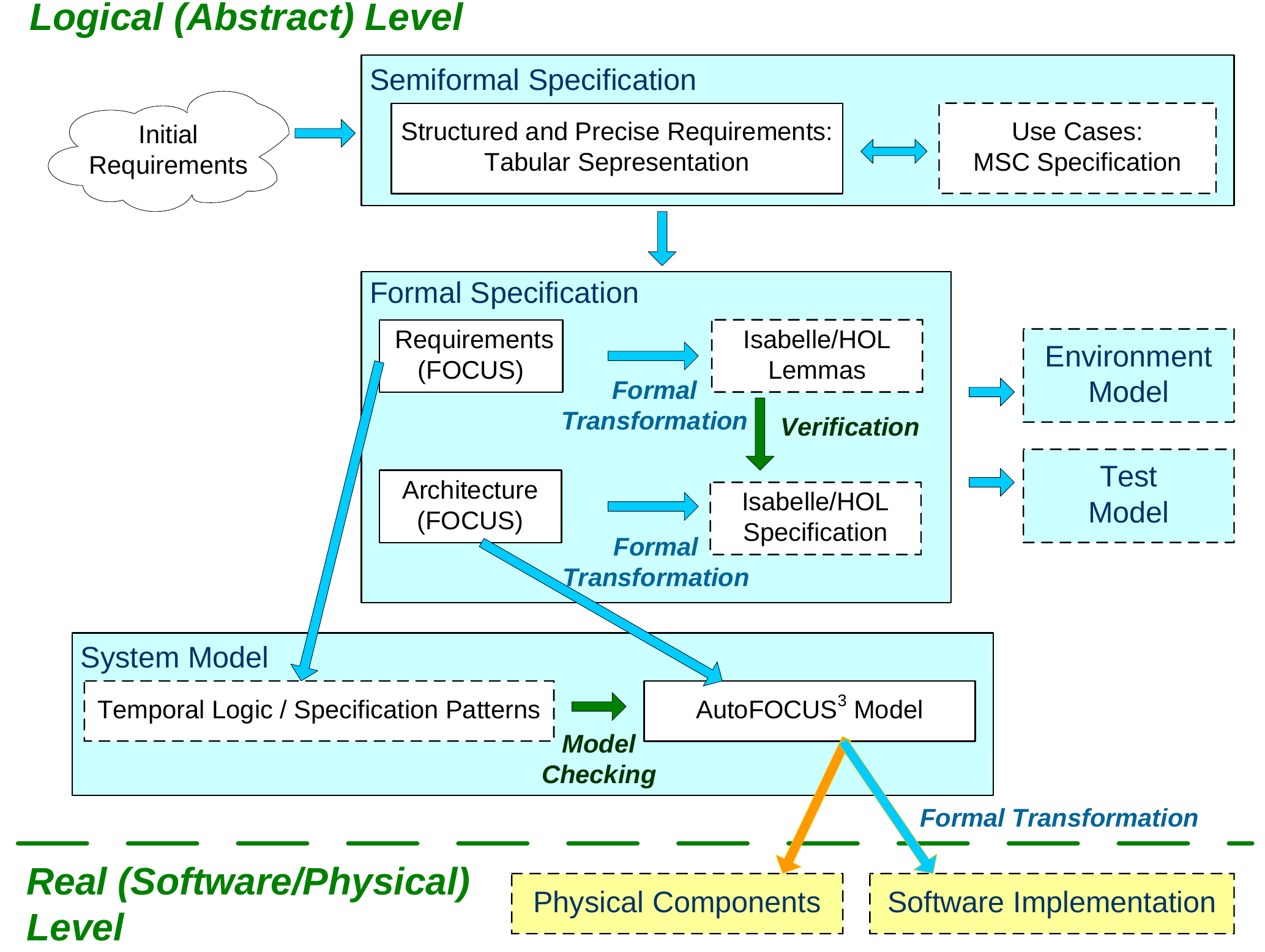}
    \label{fig:generalMethodology}
    \caption{Generalized Development Methodology}
  \end{center}
\end{figure}

Fig.~\ref{fig:generalMethodology} illustrates the structure of the generalized development methodology in a top-down manner, however, developing a large system a number of iterations is needed to cope with gain of knowledge about the system, as well as modifications in the requirements.
 The boxes represent here the artifacts that have been developed and the arrows show from which other artifacts they were derived. 
The process starts by structuring of initial requirements   
  in the way they follow specific syntactic patterns:  
  this first step raises the level of precision by transforming the free text requirements into a structured form using specific pre-defined syntactic patterns as presented in~\cite{fleischmann08}. 
An informal specification consists of a set of words, which can be distinguished into two categories: content words and keywords.
Content words are system-specific words or phrases, e.g., \emph{``Off-button is pressed''}. The set of all content words forms the \emph{logical interface} of the system, which can be understood as some kind of domain specific glossary that must be defined in addition. 
Keywords are domain-independent and  form relationships between the content words (e.g., ``if'', ``then''). 
Thus, a semiformal specification consists of a number of requirements described via  textual pattern, which is easily be understood also by engineers unfamiliar with the formal methods. 
Using this description to structure  the informal specification, we can find out missing information quite fast. 
Furthermore, we identify possible synonyms that must be unified before
proceeding to a formal specification.  
This specification can be  schematically  rewritten to a \emph{Message Sequence Charts} (MSCs, cf.~\cite{MSC:Kluwer2003}) 
representation, as an optional step relevant for  highly interacting systems.

The methodology proceeds by the translation of semiformal specification to \Focus~\cite{focus}, a framework for formal specifications
and development of distributed interactive systems, preferred here over other specification frameworks, since it 
has an integrated notion of time and  
provides a number of specification techniques for distributed systems and concepts of refinement, as well as graphical notation, which is extremely important when we are dealing with systems of industrial size. 
We represent in  \Focus  two kinds of specifications: 
system requirements and  architecture.  
This  prepares the ground to verify the system architecture  
against the requirements by translating both 
to the theorem prover Isabelle/HOL~\cite{npw} via 
the framework ``\Focus on Isabelle''~\cite{spichkova}.  
As the next step, we translate the architecture specification 
to a representation in the CASE tool AutoFocus~\footnote{The \textsc{AutoFocus} homepage: \url{http://af3.in.tum.de}} 
to simulate the system. 
The requirements specification can be schematically translated to temporal logic or specification patterns, which gives basis to model-check the model (cf.~\cite{Campetelli11}). 
The integration of model checking in AutoFOCUS  approach usability at the following points: tight coupling of verification properties with model elements, different specification languages for the formulation of properties, and visualization of counterexamples. Dealing with these issues leads to one of the first model-based development environments incorporating property specification, model checking and debugging.
Optionally, we can also represent an environment or a test model if this benefit the analysis of the concrete system. 
Finally, we can switch from the logical to the real level, where we have to split our model into  software and physical components.
Transformation  to a corresponding C code can be done using the corresponding code generator: we have shown that the C program produced by our code generator is an admissible simulation of the model.

\vspace{-3mm}
\section{Related Work} 
There are many approaches on mechatronic/cyber-physical systems, however, most of them do not focus on the logical  level of the system representation and loose the advantages of the abstract respresentation: a better overview, possibility to validate the system on the earlier phases, etc. 
For instance, the work presented in~\cite{Vogel-Heuser_IECON} defines an extensive support to the components communication and time requirements, while the model discussed in~\cite{IEEE_INDIN_2011} proposes a complete model of the processes with communication. Nevertheless, in our opinion one limitation of such approaches is that the system is represented with a flat view, that is, there is only a single abstraction level for represent it. That could be a disadvantage in the project of a cyber-physical system, where experts of different domains should be able to cooperate and work to different views and abstraction levels of the system. 
Modeling theories for distributed hybrid system as SHIFT~\cite{Deshpande97shift:a} and R-Charon~\cite{KratzSPL06} guarantee a complete simulation and compilation of the models, but they have no verification support.
The same limitation is for UPPAAL~\cite{DBLP:conf/sfm/BehrmannDL04} and PHAVer~\cite{Beek06syntaxand}, which provide the simulation, but a limited verification with restricted dynamics and only for small fragments.

\vspace{-3mm}
\section{Conclusion}
In this paper we have suggested to reuse the generalization of two existing methodologies for the development of software systems to apply them within the cyber-physical
domain, according to the results of three case studies motivated and supported by DENSO Corporation and Robert Bosch GmbH. 
Until a certain abstraction level we can 
use the existing methodologies for the development of software systems also within the cyber-physical domain to benefit from the advantages these techniques have shown, as well as extend the modeling artifacts for the domain features.

\bibliographystyle{eptcs}

\end{document}